\documentclass[fleqn,usenatbib]{mnras}

\usepackage[T1]{fontenc}
\usepackage[svgnames]{xcolor}

\DeclareRobustCommand{\VAN}[3]{#2}
\let\VANthebibliography\thebibliography
\def\thebibliography{\DeclareRobustCommand{\VAN}[3]{##3}\VANthebibliography}

\usepackage{graphicx}	
\usepackage{amsmath}	
\usepackage{amssymb}	
\usepackage{color}

\newcommand{\rd}{\color{red}}

\def\noterd #1]{{\bf \rd #1]}}

\title[A Criterion for Magnetars Producing Giant Flares]{A Criterion for Magnetars Producing Giant Flares}

\author[Y. Meng et al.]{
Y. Meng, $^{1,2,3}$
Qian-Sheng Zhang,$^{1,4,5,6}$\thanks{E-mail: zqs@ynao.ac.cn (QSZ)}
and J. Lin$^{1,2,3,4}$
\\
$^{1}$Yunnan Observatories, Chinese Academy of Sciences, 396 Yangfangwang, Guandu District, Kunming 650216, China\\
$^{2}$Center for Astronomical Mega-Science, Chinese Academy of Sciences, 20A Datun Road, Chaoyang District, Beijing 100012, China\\
$^{3}$Yunnan Key Laboratory of Solar Physics and Space Science, Kunming, Yunnan 650216, China\\
$^{4}$University of Chinese Academy of Sciences, Beijing 100049, China\\
$^{5}$Key Laboratory for the Structure and Evolution of Celestial Objects, Chinese Academy of Sciences, \\
396 Yangfangwang, Guandu District, Kunming 650216, China\\
$^{6}$International Centre of Supernovae, Yunnan Key Laboratory, Kunming 650216, China
}

\date{Accepted XXX. Received YYY; in original form ZZZ}

\pubyear{2020}

\begin{document}
\label{firstpage}
\pagerange{\pageref{firstpage}--\pageref{lastpage}}
\maketitle

\begin{abstract}
In this work, the straight flux rope in the model of giant flares on magnetars (Meng et al. 2014)
was replaced with a curved one and the equilibrium behavior of the flux rope was investigated.
Two footpoints of the flux rope are anchored to the spherical surface of magnetar. The forces acting
on the flux rope include magnetic tension, magnetic pressure, curvature force, gravity. The
equilibrium in the flux rope, so as in the global configuration, is achieved as these
forces offset each other. Changes in the background environment drive the configuration to evolve
through a set of equilibria in a quasi-static fashion until the critical point is reached and the
loss of equilibrium in the configuration occurs, invoking a giant flare. We establish a criterion to
identify magnetars capable of producing giant flares. Among the four forces, the curvature force
as well as the magnetic compression tend to expel the flue rope outward. In a given magnetic configuration,
the curvature force and magnetic compression are proportional to the square of the current intensity of
the flux rope, which is determined by the frozen-flux condition and background magnetic field strength.
 We find that only when $\eta = G M m / ({{{R^4}{B_0}^2}})< 0.48 $ is satisfied, the system reaches a critical point and potentially undergoes catastrophe.
 Here, $G, M, m, R$, and $B_{0}$ are the gravitational constant, the mass of neutron star, the mass of flux rope, the radius of neutron
 star, and the surface magnetic field strength of neutron star, respectively. The physical meaning of
 this criterion is that when $\eta \propto m / B_{0}^{2}$ is small enough, the curvature force and
 magnetic pressure can be sufficiently large to overcome gravitational confinement. This criterion
 establishes a basis for identifying magnetars capable of producing giant flares.
\end{abstract}




\section{Introduction} \label{SecIntro}
Soft gamma repeaters (SGRs) and anomalous X-ray pulsars (AXPs) are widely believed to
be magnetars - a small class of spinning isolated neutron stars \citep{DT92,Kel98,TM01,Lyu03,HL06}.
The ultra-strong magnetic fields of magnetars ($\emph{B}$ $\geq$ $10^{15}$ G )
is $10^{2}\sim$ even $10^{3}$ times stronger than that of the radio pulsars.
The persistent and bursting emissions from a magnetar are powered by the
dissipation of non-potential (current-carrying) magnetic field in the magnetosphere \citep{DT92,TD96,TLK02,Lyu06}.
Extremely rarely, an SGR produces a giant flare with
enormous energy ($\backsimeq 10^{44}-10^{47}$ erg) and long duration of burst.
These exceptionally powerful outbursts consist of a precursor of 1 s, a very short ($\sim 0.1-0.2$~s)
hard spike of $\gamma$-rays emissions containing most of the flare energy and a pulsating tail emission lasting about 380
seconds \citep{Hur05}.

So far, three SGRs producing rare giant flares have been observed\citep{Maz79,Hur99,Hur05},
which are SGR 0526-66 on 5 March 1979 \citep{Maz79}, SGR 1900+14 on
27 August 1998 \citep{Hur99,Kou99,Vrb00}, and SGR 1806-20 on 27 December
2004 \citep{Hur05,Pal05}, respectively. The giant flare from SGR 1806-20 was much
more energetic than the other two events \citep{Hur05,Pal05}.
Its initial $\gamma$-ray spike released the
energy of $\sim 10^{46}$ erg within $\sim 0.2$~s, and its rising and
falling phase lasting $\tau_{rise}\leq 1$ ms and $\tau_{fall}\approx 65$ ms,
respectively. Followed by the main spike, a tail emission that modulated
at the neutron star spin period (7.56~s) lasted about 380 ~s. \citep{Hur05,Pal05}.

The energy of a magnetar eruption is widely believed to
come from the  magnetic field. Magnetic energy dissipation is commonly considered to account
for the energetic behavior \citep{DT92,WT06,Mer08}.
So far, two models of giant flares of SGRs exist, which
depend on the location of the magnetic energy storage prior to the
eruption: in the crust \citep{TD01} or in the magnetosphere space\citep{Lyu06}.
In the crust model, the eruption is driven by the motion of evolution of the crust
with the time scale of $0.2 - 0.5$s \citep{TD95}.
And in the magnetosphere model, the energy is gradually built up in the magnetosphere of the
star beforehand, and the eruption is triggered by the loss of equilibrium in the magnetic configuration
include the magnetosphere, and driven release of the stored energy.
In the environment of the magnetosphere, the time scale of the loss of equilibrium is
about 30$\mu$s \citep{Lyu03}.

Observations of the
giant flare from SGR 1806-20 on 27 December 2004 showed that it experienced a
very short rising phase, $\sim$ 0.25~ms \citep{Pal05}.
Therefore, the crust model is difficult in explaining the very short rising time of
giant flares \citep{LL12,Link14} and the magnetospheric model is more appropriate to account for the
dynamic of magnetar's giant flares \citep{MLZ14,HY14b}.
In the magnetosphere model, the magnetic energy is slowly
stored in the magnetosphere on time scales
much longer than that of the giant flare itself, until the
system reaches a critical point at which the equilibrium becomes unstable.
Then further evolution in the system occurs and leads to
the eruption similar to solar flares and coronal mass ejections (CMEs)
taking place in the solar atmosphere \citep{Lyu06,MLZ14}, which suggests that
solar flares and CMEs provide an important references for magnetar giant flares.

\citet{Mas10} used a magnetic reconnection model of the solar flare proposed by \citet{SY99} to explain
 magnetar giant flares with taking account of chromospheric evaporation.
 In their work, the preflare activity produces a
baryon-rich prominence. Then the prominence erupts as a result of
magnetic reconnection to produce a giant flare, and the eruption constitutes
the origin of the observed radio-emitting ejecta
associated with the giant flare from SGR 1806-20
\citep{Tay05,Cam05,Gae05,Mas10}. By analogy with the solar CME events,
\citet{MLZ14} developed an MHD  model for magnetar giant flares via an analytic approach.
In their model, the rotation and/or displacement of the crust causes the magnetic field to twist
and deform, leading to the formation flux rope in the
magnetosphere and the energy accumulation in the system. When the energy and the helicity stored in
the configuration reach a threshold, the system loses its equilibrium, the flux rope is ejected outward in a catastrophic
way, and magnetic reconnection helps the catastrophe develop to a plausible eruption.
According to the model, they obtained a light curve that agreed with observations.

MHD numerical modeling has significantly contributed to elucidating the eruption processes in magnetars.
\citet{Pfr12a,Pfr12b,Pfr13} demonstrated that crustal shearing drives magnetospheric expansion and reconnection,
linking magnetic twist evolution to observed flares.
Building upon solar CME theory and model, the accumulation and release of magnetic flux rope energy within the magnetosphere,
along with the process of magnetic field reconfiguration in dipole/multipole field, have been extensively investigated \citep{Yu11,Yu12,HY14a,HY14b,Sha23}.
Recent 3D force-free simulations have significantly advanced our understanding of magnetar outburst mechanisms.
By modeling 3D force-free twisted magnetospheres of neutron stars, magnetar outbursts could be triggered through
the gradual energy accumulation, perturbations, and sudden magnetic reconfiguration. \citep{Car19,Mah19,Mah23,Yua22,Ste23,Rug24}.

A similar scenario or approach to that of \citet{Mah23} is also observed or applied in the field of solar physics in both observations \citep{Yan18} and numerical simulations \citep{Jel21}. In the solar case, how much twist is needed to trigger the eruption is still an open question. Numerical experiments showed that the eruption occurs quickly after the helical structure forms in the relevant configuration \citep[e.g., see also][]{MK94}. On the other hand, eruptions may occur easily in the configuration including a pre-existing flux rope filled with entangled and braided magnetic field \citep[see also][]{Lel03,Fel06}.

The initiation and development of magnetar giant flares have been
extensively studied, several models have been suggested \citep{TD95,TD01,Lyu06,
Mas10,Yu11,Yu12,HY14a,HY14b,MLZ14},and the related numerical simulations have been carried out as well.
But the origin and development of giant flares remains unclear. The detailed physical process of the
magnetic energy storage and release, as well as which kind of magnetar could produce giant flare are
still an open question. In this work, we construct a
flux rope model for magnetar giant flares on the basis of the magnetohydrodynamical (MHD) model developed by \citet{MLZ14},
and the CME models developed by \citet{Ltl98} in Section \ref{SecModel}, and study the dynamical process of energy storage and release
of the giant flares in Section \ref{SecRes}. We discuss these results  and deduced the criterion for magnetar
that could produce giant flares in Section \ref{SecDis}.
Finally, we summarize this work in Section \ref{SecCon}.

\section{Constitution of the Model} \label{SecModel}

In the framework of the catastrophe model of solar eruptions,
the evolution in the system includes two stages and
a triggering process.
In the first stage, the magnetic energy is gradually
stored in the coronal magnetic field in response to the
change in the background magnetic field, such as the motion
of the footpoint of the coronal magnetic field anchored in the
magnetosphere, and the change in the background field.
In the second stage, the stored energy is quickly released.
The two stage connect to each other by a triggering process
behaving as the loss of equilibrium in the configuration.
This process is also known as the catastrophe, and
could be either ideal or non-ideal MHD depending on the fashion
of the system evolution or the detailed structure in the magnetic field involved
\citep[e.g., see discussions of][]{FI91,Lel03}.
(Note: "ideal" here means no resistivity, and "non-ideal" means that the resistivity is included.)
But magnetic reconnection is crucial for the energy release to produce
the flare \citep{Par63,Pet64} and to allow the loss of equilibrium to
develop to a plausible eruption smoothly \citep{LF00,Lin02}.

We consider a slow evolution through a series of quasi-static equilibria,
where changes occur over a time interval that is longer than the characteristic
time of the system. In an environment filled with magnetized plasma, this
characteristic time scale is the Alfv\'{e}n time scale. In the present work,
the evolution in the magnetic configuration of interest is slow prior
to the giant flare, and it is the behavior of the magnetic configuration in such slow
evolution that constitutes the focus of our study.

\subsection{Model Description}
In the case of the giant flare on the magnetar, the changes in the
background field could be caused by the cracking and the motion of the crust.
It is believed that crust cracking could take place on the magnetar
from time to time \citep[see][]{Rud91a,Rud91b,Rud91c}.
The surface magnetic field of a spin-down crust-cracking neutron star might break
into large surface patches\citep{Rud91c}, meanwhile the Lorentz force due to the strong
magnetic field acting on the crust could result in the build up of stress, and
further causes pieces of the broken crust to rotate on the equipotential surface until the
stress acting on the lattice exceeds a critical value \citep{Lyu06}.
This processes could also lead to the storage of the magnetic stress and the energy in the magnetosphere.
Therefore, the energy driving the giant flare on the magnetar is transported
from the inside of the magnetar via a reasonable mechanism, and then stored in the
magnetosphere before the eruption although details of the energy storage accounting on the magnetar is
somewhat different from that on the Sun.

By analogy with the CME events on the Sun, \citet{MLZ14}
developd a theoretical model for magnetar giant flares via an analytic approach
on the basis of Lin-Forbes model \citep{LF00}.
In the model of \citet{MLZ14}, the rotation and/or displacement of the crust
causes the field to twist and deform, leading to flux rope formation in the
magnetosphere and energy accumulation in the
related configuration. When the energy and helicity stored in the configuration
reach the threshold, the system loses its equilibrium,
the flux rope is ejected outward in a catastrophic way, and magnetic reconnection
helps the catastrophe develop to a plausible eruption.
The released free magnetic energy is converted into radiative energy, kinetic energy
and gravitational energy of the flux rope. They calculated the light curves of
the eruptive processes for the giant flares of SGR 1806-20, SGR 0526-66 and
SGR 1900+14, which are in good agreement with the observed light curves
of giant flares.

Although the model succeeds in explaining observations on some aspects,
it did not specifically look into physical details of the evolution of
the equilibrium in the magnetic configuration, which is related to fact
that the magnetar has a spherical surface, to which the two ends of the
cowed flux rope anchored, a curvature force acting on the cowed flux
rope \citep[e.g., see also][]{Ltl98,Lin02}, and magnetar rotates fast.
So the model of \citet{MLZ14} can be considered simple, and it is
necessary to establish a theoretical model that is consistent with the
situation of magnetar in reality. Following the practice of \citet{Ltl98} and
\citet{Lin02}, a three-dimensional model of giant flare is developed in
the present work.

In the model of \citet{Ltl98}, the flux rope is a
torus encircling the Sun and system is axialiy symmetric. Initially,
the flux rope is suspended in the corona as a result of the  balance
among magnetic tension, compression, and curvature forces, and the
balance is  eventually lost as the photospheric source of the coronal
field slowly decay with time. The evolution of the system shows
catastrophic behavior and the flux rope of large radius behaves more
energetic than that of smaller radius. The loss of equilibrium could
lead to the rapid formation of a current sheet, and fast magnetic
reconnection occurring in the current sheet would help the flux rope
escape smoothly. \citet{Lin02} further looked into the evolution behavior
of the magnetic configuration including a flux rope of half circle with
two ends anchored to the solar surface, which is modeled by a plane.
The gradual evolution in the configuration in response to the slow
decay of the background magnetic field eventually turns into dynamic
evolution after the loss of equilibrium of the flux rope via catastrophe.
Combining the physical scenerios and techniques used to deal with
mathematics occurring in the above works, we construct a three-dimensional
MHD model for the giant flare on magnetars, which includes a partly
circular flux rope with two ends anchored to the spherical surface of
magnetars. The related mathematics is given in next section.

\subsection{Formulization of the Model}
In the calculations below, we assume that the mangetar is a sphere.
Owing to the scarcity of observational data currently, very little
is known about the true shape of the magnetar. So, at least, as the
zeroth approximation, the magnetar is treated as a sphere.

We set the stellar radius as unity and look into the evolutionary
behaviors of the flux rope of the partly circle shape. The vector
potential satisfies the Poisson equation:
\begin{eqnarray} \label{AJbasic}
{\nabla ^2}{{\bf{A}}}  =  - \frac{{4\pi }}{c}{{\bf{J}}} ,
\end{eqnarray}
where ${\bf{J}}$ is the electric current density in the flux rope, and $c$
is the light speed. We use the Gaussian unit system in this work. The boundary
conditions in our model are that ${\bf{A}}$ is the background field ${\bf{A}}_0$
at the stellar surface and ${\bf{A}}$ vanishing at infinity.
Outside the stellar surface (i.e., $ r \geq 1 $), ${\bf{A}}$ can be obtained by solving the
above equation using the Green's function method \citep[e.g.,][]{Has13} (see Appendix A for details) as
\begin{eqnarray} \label{AGreen0}
&& {\bf{A}} ({\bf{r}}) = \frac{{4\pi }}{c}\int {{\bf{J}} ({\bf{r}}'){\widetilde G}({\bf{r}} ,{\bf{r}}')dV'}  \\ \nonumber
&& + \oint {_{r'=1}[{\widetilde G}({\bf{r}} ,{\bf{r}}') {\nabla' {\bf{A}} ({\bf{r}}')} - {\bf{A}} ({\bf{r}}'){\nabla' {\widetilde G}({\bf{r}},{\bf{r}}')}]dS'},
\end{eqnarray}
with the Green's function ${\widetilde G}$ satisfying:
\begin{eqnarray} \label{AGreen1}
&& {\nabla ^2}{\widetilde G}({\bf{r}},{\bf{r}}') =  - \delta ({\bf{r}}-{\bf{r}}'), \\ \nonumber
&& {\rm for~either~} r^{\prime} = 1 {~ \rm or ~} r = 1,  {\widetilde G}({\bf{r}},{\bf{r}}') = 0.
\end{eqnarray}

Using the method of image, the solution for the Green's function ${\widetilde G}$ in the above equation is
\begin{eqnarray} \label{GreenDef}
&&{\widetilde G}({\bf{r}} ,{\bf{r}}') = \frac{1}{{4\pi }}[{g_1}({\bf{r}} ,{\bf{r}}') - {g_2}({\bf{r}} ,{\bf{r}}')], \\ \nonumber
&&{g_1}({\bf{r}} ,{\bf{r}}') = \frac{1}{{\left| {{\bf{r}}  - {\bf{r}}'} \right|}}, \\ \nonumber
&&{g_2}({\bf{r}} ,{\bf{r}}') = \frac{1}{{\left| {r'{\bf{r}}  - {\bf{r}}'/r'} \right|}} = {g_1}(r'{\bf{r}} ,{\bf{r}}'/r').
\end{eqnarray}

 Therefore the solution for ${\bf{A}}$ is
\begin{eqnarray} \label{AGreen}
&&{{\bf{A}}} (r,\theta ,\phi ) = \frac{{4\pi }}{c}\int {{{\bf{J}}} ({{\bf{r}}}') {\widetilde G}({{\bf{r}}},{{\bf{r}}}') dV'} \\ \nonumber
&&+ \oint {[-{{\bf{A}}}_0(r'=1,\theta ',\phi ')]} {\left. {{\nabla' {\widetilde G}({{\bf{r}}},{{\bf{r}}}')}} \right|_{r' = 1}}dS',
\end{eqnarray}

In the case of current-free configuration ${\bf{J}} = \bf{0}$, the above equation gives the background field
${\bf{A}}_0$, which is the second term at the right side of Eq.(\ref{AGreen}).
We now focus on the first term that is denoted as ${\bf{A}}_V$, i.e., the
volume integral which is produced by the combination of the current in the
flux rope (namely $g_1$ in ${\widetilde G}$) and an effective image current
in the star (namely $g_2$ in ${\widetilde G}$). Then ${{\bf{A}}}={\bf{A}}_V+{\bf{A}}_0$.

Therefore Eq.(\ref{AGreen}) shows that
for a given section of the rope, $dV'$ is the volume element, $d{\bf{S}}'$
is the area element on the flux rope cross-section. Following the practice of
\citet{FI91} and \citet{Iea93}, the radius of the flux rope is small compared
to the scale of the global configuration to the first order approximation.

\begin{eqnarray} \label{AV}
&&{\bf{A}}_V  = \frac{{4\pi }}{c}\int {{\bf{J}} ({\bf{r}}'){\widetilde G}({\bf{r}} ,{\bf{r}}')dV'} \\ \nonumber
&& = \frac{{4\pi }}{c}\int\limits_V^{} {{\bf{J}} ({\bf{r}}'){\widetilde G}({\bf{r}} ,{\bf{r}}')(d{\bf{S}}' \cdot d{\bf{l}}' )} \\ \nonumber
&& = \frac{{4\pi }}{c}\int\limits_V^{} {{\widetilde G}({\bf{r}} ,{\bf{r}}')[{\bf{J}} ({\bf{r}}') \cdot d{\bf{S}}' ] d{\bf{l}}' } \\ \nonumber
&& = \frac{{4\pi }}{c}\int\limits_{}^{} {d{\bf{l}}'} \int\limits_S^{} {{\widetilde G}({\bf{r}} ,{\bf{r}}'){\bf{J}} ({\bf{r}}') \cdot d{\bf{S}}' } \\ \nonumber
&& = \frac{{4\pi I}}{c}\int {{\widetilde G}({\bf{r}} ,{\bf{r}}')d{\bf{l}}' },
\end{eqnarray}
where $I$ is the total electric current intensity flowing through the flux
rope and $d{\bf{l}}'$ is the length element along the flux rope and in the
direction of ${\bf{J}}$. We evaluate each component of ${\bf{A}}_V$ in
the Cartesian coordinate system, separately:
\begin{eqnarray} \label{AVxyz}
&&(A_{V,x},A_{V,y},A_{V,z}) = \frac{{4\pi I}}{c}\int {{\widetilde G}({\bf{r}} ,{\bf{r}}')(dx',dy',dz')} \\ \nonumber
&& = \frac{I}{c}\int {({g_1} - {g_2})(dx',dy',dz')}.
\end{eqnarray}

Then the magnetic field produced by the electric current inside the the flux
rope as well as the image current is
${\bf{B}}_V=\nabla  \times {\bf{A}}_V$. In the Cartesian coordinate system, since
\begin{eqnarray} \label{formulag}
&& {g_1} = \frac{1}{{\sqrt {{{(x - x')}^2} + {{(y - y')}^2} + {{(z - z')}^2}} }} \\ \nonumber
&& {g_2} = \frac{1}{{\sqrt {{{(r'x - x'/r')}^2} + {{(r'y - y'/r')}^2} + {{(r'z - z'/r')}^2}} }},
\end{eqnarray}
gradients of $g_1$ and $g_2$ are:
\begin{eqnarray} \label{grdg}
&&\frac{{\partial {g_1}}}{{\partial (x,y,z)}} = \frac{{g_1}^3}{{ - 2}}\frac{{\partial {g_1}^{ - 2}}}{{\partial (x,y,z)}} =  - (x - x',y - y',z - z'){g_1}^3 \\ \nonumber
&& \frac{{\partial {g_2}}}{{\partial (x,y,z)}} = \frac{{g_2}^3}{{ - 2}}\frac{{\partial {g_2}^{ - 2}}}{{\partial (x,y,z)}} =  - (r{'^2}x - x',r{'^2}y - y',r{'^2}z - z'){g_2}^3.
\end{eqnarray}
Therefore the components of ${\bf{B}}_V = \nabla  \times {\bf{A}}_V $ are:
\begin{eqnarray} \label{BVxyz}
&&{B_{V,x}} = \frac{I}{c}\int {} \{ [(z - z')dy' - (y - y')dz']{g_1}^3 \\ \nonumber
&& - [(r{'^2}z - z')dy' - (r{'^2}y - y')dz']g_2^3\}   \\ \nonumber
&&{B_{V,y}} = \frac{I}{c}\int {} \{ [(x - x')dz' - (z - z')dx']{g_1}^3 \\ \nonumber
&& - [(r{'^2}x - x')dz' - (r{'^2}z - z')dx']g_2^3\}  \\ \nonumber
&&{B_{V,z}} = \frac{I}{c}\int {} \{ [(y - y')dx' - (x - x')dy']{g_1}^3 \\ \nonumber
&& - [(r{'^2}y - y')dx' - (r{'^2}x - x')dy']g_2^3\}.
\end{eqnarray}

\begin{figure}
	\includegraphics[width=1.0\columnwidth]{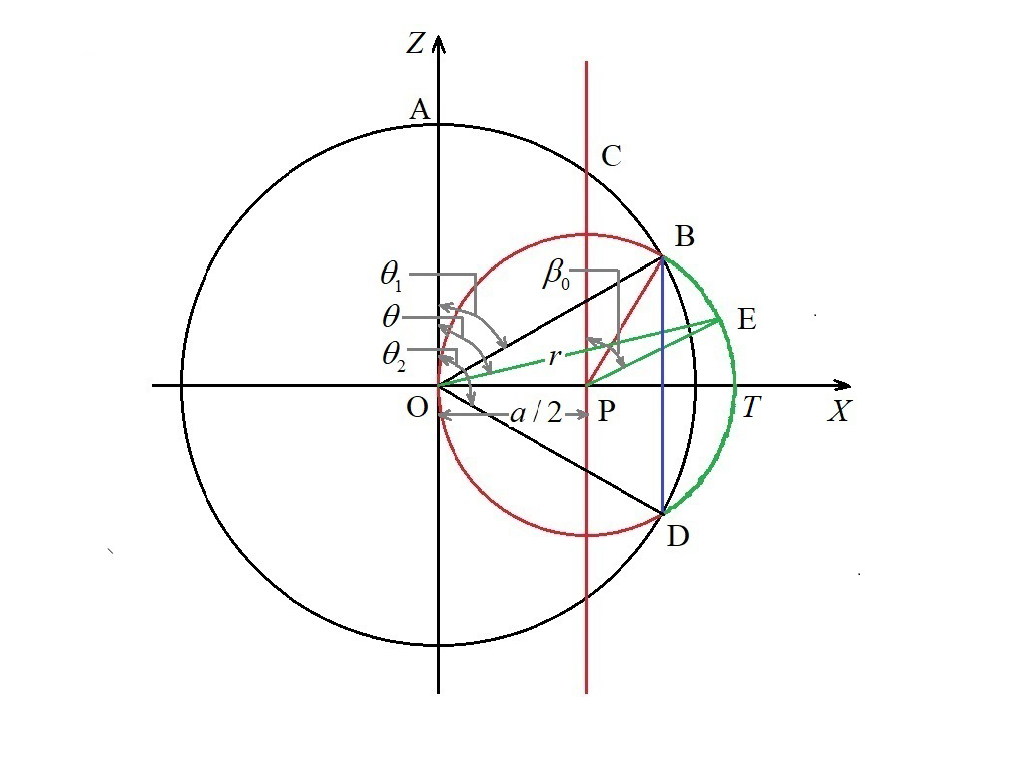}
    \caption{A sketch of the configuration of the flux rope. The star is the circle O with radius OA defined as unity. The flux rope is the $r>1$ part (green arc BED) of a circle (red) P passing the stellar center O.\textbf{ ${\rm \mid OP \mid} =a/2$, $\theta=\angle{\rm AOE}$.} The plane is the $\phi=0$ plane in the spherical coordinates or the X-O-Z plane in the corresponding Cartesian coordinate system. OA is the pole or the Z-axis. The blue line BD is the effective image current in the star. Noting that the coordinate systems are established by let the stellar center be the original and the plane contains the flux rope be the $\phi=0$ and $y=0$ plane. In the most favorable condition for the instability of the magnetic flux rope, the magnetic axis of the background dipole is perpendicular to the plane of the paper and downward passes through the origin O. The top point of the rope is denoted as point T with the coordinates $x=a$ and $z=0$. }
    \label{ropesketch}
\end{figure}

We assume that the flux rope is initially located on a plane passing the
stellar center. A spherical coordinates system and the corresponding Cartesian coordinate
system can be established by setting $\phi=0$ and $y=0$ on the plane. We further assume that the flux rope is the part of a circle of $r>1$
with a diameter of $a$, and the stellar center is located on the plane of the
circle (see Fig.\ref{ropesketch}). The equation of the flux rope in the corresponding Cartesian coordinate
system is:
\begin{eqnarray} \label{0rope}
x^{2} + z^{2} - \frac{a}{2} x = 0,
\end{eqnarray}
Substituting $x = r \sin \theta$ and $z = r \cos \theta$ into the above equation gives:
\begin{eqnarray} \label{rope}
r = a\sin \theta, \{ {\theta _1} \le \theta \le {\theta _2}  \},
\end{eqnarray}
where ${\theta _1 = \arcsin (a^{-1})}$ and ${\theta _2 = \pi  - {\theta _1}}$ are the values of ${\theta}$ at the two
footpoints of the flux rope on the star surface, respectively.
On the rope, since $x = r \sin \theta = a{{\sin }^2}\theta$ and $z = r \cos \theta = a \sin\theta \cos\theta$,
the differentials with respect to $x$ and $z$ are:
\begin{eqnarray} \label{rope}
&&dx = a\sin 2\theta d\theta,  \\ \nonumber
&&dz = a\cos 2\theta d\theta.
\end{eqnarray}
Substituting the differentials of $x'$ and $z'$ on the rope (i.e., $dx' = a\sin 2\theta' d\theta'$ and $dz' = a\cos 2\theta' d\theta'$) into Eq.(\ref{AVxyz}) \& Eq.(\ref{BVxyz}) yields:
\begin{eqnarray} \label{ropeAVxyz}
&&{A_{V,x}} = \frac{aI}{c}\int\limits_{{\theta _1}}^{{\theta _2}} {({g_1} - {g_2})(2\sin \theta '\cos \theta ')d\theta '},  \\ \nonumber
&&{A_{V,y}} = 0, \\ \nonumber
&&{A_{V,z}} = \frac{aI}{c}\int\limits_{{\theta _1}}^{{\theta _2}} {({g_1} - {g_2})(1 - 2{{\sin }^2}\theta ')d\theta '}.  \\ \nonumber
 \end{eqnarray}
and
\begin{eqnarray} \label{ropeBVxyz}
&&{B_{V,x}} = \frac{aIy}{c}\int\limits_{{\theta _1}}^{{\theta _2}} [(2{{\sin }^2}\theta ' - 1){g_1}^3 \\ \nonumber
&& - {a^2}(2{{\sin }^4}\theta ' - {{\sin }^2}\theta '){g_2}^3 ] d\theta ',  \\ \nonumber
&&{B_{V,y}} = \frac{aI}{c}\int\limits_{{\theta _1}}^{{\theta _2}} {\{ [x + (a - 2x){{\sin }^2}\theta ' - 2z\sin \theta '\cos \theta ']{g_1}^3} \\ \nonumber
&& + a[2a(z{\sin ^3}\theta '\cos \theta ' + x{\sin ^4}\theta ') - (1 + ax){\sin ^2}\theta ']{g_2}^3\} d\theta ', \\ \nonumber
&&{B_{V,z}} = \frac{aIy}{c}\int\limits_{{\theta _1}}^{{\theta _2}} {(2\sin \theta '\cos \theta '{g_1}^3 - 2{a^2}{{\sin }^3}\theta '\cos \theta '{g_2}^3)d\theta '}.
\end{eqnarray}
In the above integrals, the Green's functions $g_1$ and $g_2$ in the spherical coordinates are:
\begin{eqnarray} \label{defpq}
&&g_1 = \frac{1}{{\sqrt {{r^2} + {a^2}{{\sin }^2}\theta ' - 2ar\sin \theta '\cos \gamma } }}, \\ \nonumber
&&g_2 = \frac{1}{{\sqrt {{a^2}{r^2}{{\sin }^2}\theta ' + 1 - 2ar\sin \theta '\cos \gamma } }},
\end{eqnarray}
where
\begin{eqnarray} \label{defgamma}
 \cos \gamma  = \cos \theta \cos \theta ' + \cos \phi \sin \theta \sin \theta '.
\end{eqnarray}

Define symbols $p_{smn}$ and $q_{smn}$ as:
\begin{eqnarray} \label{defpq}
&&{p_{smn}} := \int\limits_{{\theta _1}}^{{\theta _2}} {g{{_1}^s}{{\sin }^m}\theta '{{\cos }^n}\theta 'd\theta '} , \\ \nonumber
&&{q_{smn}} := \int\limits_{{\theta _1}}^{{\theta _2}} {g{{_2}^s}{{\sin }^m}\theta '{{\cos }^n}\theta 'd\theta '},
\end{eqnarray}
and then the components of ${\bf{A}}_V$ and ${\bf{B}}_V$ in Eqs.(\ref{ropeAVxyz})
and (\ref{ropeBVxyz}) become:
\begin{eqnarray} \label{defpq}
&&{A_{V,x}} = \frac{{aI}}{c}[2({p_{111}} - 2{q_{111}})] \\ \nonumber
&&{A_{V,z}} = \frac{{aI}}{c}[({p_{100}} - {q_{100}}) - 2({p_{120}} - {q_{120}})] \\ \nonumber
&&{B_{V,x}} = \frac{{aIy}}{c}[(2{p_{320}} - {p_{300}}) - {a^2}(2{q_{340}} - {q_{320}})] \\ \nonumber
&&{B_{V,z}} = \frac{{aIy}}{c}(2{p_{311}} - 2{a^2}{q_{331}}) \\ \nonumber
&&{B_{V,y}} = \frac{{aI}}{c}\{ [(a - 2x){p_{320}} - 2z{p_{311}} + x{p_{300}}] \\ \nonumber
&& + a[2a(x{q_{340}} + z{q_{331}}) - (ax + 1){q_{320}}]\}.
\end{eqnarray}

In the above equations, the $p$-terms are contributed by the internal current
of the flux rope and the $q$-terms are contributed by the image current.
Calculations for each $p_{smn}$ were given in details in Appendix B. The
results for $q_{smn}$ were not given specifically, but it is trivial to obtain
them by replacing $r$ with 1 and replacing $a$ with $ar$, respectively, in
the results for $p_{smn}$. With $p_{smn}$ and $q_{smn}$ being known, ${\bf{A}}_V$
and ${\bf{B}}_V$ are then deduced from Eqs.(\ref{defpq}).

\subsection{Equations Governing the motion of the flux rope} \label{SecEquation}

Figure \ref{ropesketch} describes the surface of the star, and the location and
the shape of the flux rope. The star surface is shown as the circle with
radius of OA ($ {\rm \mid OA \mid} =1$). The stellar mass is set to be unity,
the mass of the rope is $m$ and the density of the rope is assumed homogeneous,
and the arc BED is the flux rope. We now evaluate forces acting on a line
element of the flux rope at an arbitrary point E. The polar angle of point E is
$\theta=\angle{\rm AOE}$, define $\beta_{0}=\angle{\rm CPE}$ and then $\beta_{0}=2\theta-\pi/2$,
and the line element $d{\bf{l}}$ of the flux rope at point E is:
\begin{eqnarray} \label{lineelement}
d{\bf{l}} = \frac{a}{2} d\beta_{0} (\cos \beta_{0} {\bf{i}} - \sin \beta_{0} {\bf{k}}) = a d\theta (\sin 2\theta {\bf{i}} + \cos 2\theta {\bf{k}}).
\end{eqnarray}
Here $\bf{i}$ and $\bf{k}$ are the unit vectors in x- and z-directions,
respectively. The total force on the line element is
\begin{eqnarray} \label{forcelineelement}
d{\bf{f}} = \frac{I}{c}{B_s}dl \frac{{2\bf{r}}-{\bf{i}}a}{|{2\bf{r}}-{\bf{i}}a|} + \frac{I}{c}d{\bf{l}} \times {{\bf{B}}_{e}} - \frac{{G{\bf{r}}}}{{{r^3}}}dm,
\end{eqnarray}
where the terms on the right hand side from left to right are the curvature
force, the magnetic force and the gravity, respectively, with $G$ being the
gravitational constant. The external magnetic
field at the axis of the flux rope, ${\bf{B}}_{e}$, includes the background
field ${\bf{B}}_0$ and the magnetic field ${\bf{B}}_q$ caused by image current,
where the components of ${\bf{B}}_q$ are obtained from Eq.(\ref{defpq}) by
setting $p$ terms equal to zero on the flux rope, ${\bf{B}}_q = B_{q,y}{\bf{j}}$,
and $B_s$ related to the curvature force is \citep{Shafranov1966}

\begin{eqnarray} \label{formulabs}
{B_s} = \frac{I}{{ca}}\left[\ln \frac{{8a}}{{{r_0(\theta)}}} - 1 \right],
\end{eqnarray}
where $r_0$ is the radius of the flux rope cross section that is determined
by the internal equilibrium of the flux rope such that $ r_0 I = const$, \citep[e.g., see Appendix C in][for details]{Ltl98}.

This curvature force is the radial self-force acting on the current,
it is caused by the compression of the poloidal field along the inside edge
of the torus, and was evaluated by \citet{Shafranov1966}. Its magnitude is
equal to the current, $I$, times the magnetic field $B_s$, where $B_s$ is the
field created by a ring current, $I$, distributing over its circular cross
section of radius $r_0$.

We note here the centrifugal force does not appear in Eq.(\ref{forcelineelement}),
which does not seem consistent with the purpose of this work that is to
looking into the impact of the rotation of magnetar the occurrence of the
giant flare. This could be justified in this way. The centrifugal force
itself acting on the flux rope is small compared to those showing in Eq.(\ref{forcelineelement}).
If a star rotates so fast that the rotating speed at its equator on the
surface equals to its first cosmic velocity, the mass at the equator will
escape from the star. For a typical magnetar, its mass $M=3\times10^{33}$g
and $R=10^6$cm, so its first cosmic velocity $v=\sqrt{GM/R}$, and the
corresponding period $\tau=2 \pi R/V = 2 \pi R (GM/R)^{-1/2}= 2 \pi(R^3/GM)^{1/2}=0.4$~ms.
This period is even short than that of the millisecond pulsar, and implies
that the centrifugal force itself acting on the mass above the surface of
the magnetar, of which the rotation period is around 1s, could be ignored.
In fact, the impact of the rotation of magnetar on the behavior of the
magnetic structure and the associated mass in the magnetosphere embodies
in the constraints of the rotation on the magnetar magnetic field intrinsically.
We shall discuss this issue later.

Taking the geometry of the rope into account, noting $dl=a d\theta$,
$l=a(\theta_2-\theta_1)$ and $dm=mdl/l$, the total force on the flux rope given
in Eq.(\ref{forcelineelement}) becomes:
\begin{eqnarray} \label{forcetotal}
&&{\bf{f}} = \int\limits_{{\theta _1}}^{{\theta _2}} {} \{ \frac{{aI}}{c}[({B_s} + {B_{e,y}})( - \cos 2\theta {\bf{i}} + \sin 2\theta {\bf{k}})  \\ \nonumber
&& + ({B_{e,x}}\cos 2\theta  - {B_{e,z}}\sin 2\theta ){\bf{j}}] - \frac{m}{{{\theta _2} - {\theta _1}}}\frac{{G{\bf{r}}}}{{{r^3}}} \} {d\theta }.
\end{eqnarray}

Here $\bf{j}$ is the unit vector in the y-direction. The background magnetic
field ${\bf{B}}_0$ is a dipolar one with the axis of the dipole being
${\bf{k'}} = {\alpha} {\bf{i}} + {\beta} {\bf{j}} + {\lambda} {\bf{k}}$,
where ${\alpha}$, ${\beta}$ and ${\lambda}$ are free parameter satisfying
${\alpha}^2+{\beta}^2+{\lambda}^2=1$, therefore ${\bf{A}}_0$ and ${\bf{B}}_0$ are
\begin{eqnarray} \label{B0vector}
&&{\bf{A}_0}(r) = \frac{{u{B_0}}}{{{r^3}}}\bf{k}' \times \bf{r}, \\ \nonumber
&&{{\bf{B}}_0}({\bf{r}}) = u{B_0}\frac{{3{\bf{k}}' \cdot {\bf{r}}{\bf{r}} - {r^2}{\bf{k}}'}}{{{r^5}}} \\ \nonumber
&& = \frac{{u{B_0}}}{{{r^5}}}[3(\alpha x + \beta y + \lambda z){\bf{r}} - {r^2}{\bf{k}}'],
\end{eqnarray}
where $B_0$ is magnetic field on the stellar surface at the magnetic equator
in quiescence, and $u$, which slowly decreases with time, is the magnetic field
strength relative to the quiescent one. Components of ${\bf{B}}_0$ are worked
out as
\begin{eqnarray} \label{B0xyz}
&&{B_{0,x}} = {{\bf{B}}_0} \cdot {\bf{i}} = \frac{{u{B_0}}}{{{r^5}}}[3(\alpha x + \beta y + \lambda z)x - \alpha {r^2}], \\ \nonumber
&&{B_{0,y}} = {{\bf{B}}_0} \cdot {\bf{j}} = \frac{{u{B_0}}}{{{r^5}}}[3(\alpha x + \beta y + \lambda z)y - \beta {r^2}], \\ \nonumber
&&{B_{0,z}} = {{\bf{B}}_0} \cdot {\bf{k}} = \frac{{u{B_0}}}{{{r^5}}}[3(\alpha x + \beta y + \lambda z)z - \lambda {r^2}].
\end{eqnarray}

Substituting ${\bf{B}}_0$ into Eq.(\ref{forcetotal}) gives the total force on
the flux rope
\begin{eqnarray} \label{fxyz}
&&{f_x}=\int\limits_{{\theta _1}}^{{\theta _2}} \{ - \frac{{aI}}{c}\left({B_s} + {B_{q,y}} - \frac{{\beta u{B_0}}}{{{r^3}}}\right) \cos 2\theta \\ \nonumber
&&- \frac{m}{{{\theta _2} - {\theta _1}}}\frac{{Gx}}{{{r^3}}} \} d\theta , \\ \nonumber
&&{f_y}=\frac{{aI}}{c}\int\limits_{{\theta _1}}^{{\theta _2}} {({B_{0,x}}\cos 2\theta  - {B_{0,z}}\sin 2\theta )d\theta } , \\ \nonumber
&&{f_z}=0.
\end{eqnarray}

\begin{figure}
	\includegraphics[width=1.0\columnwidth]{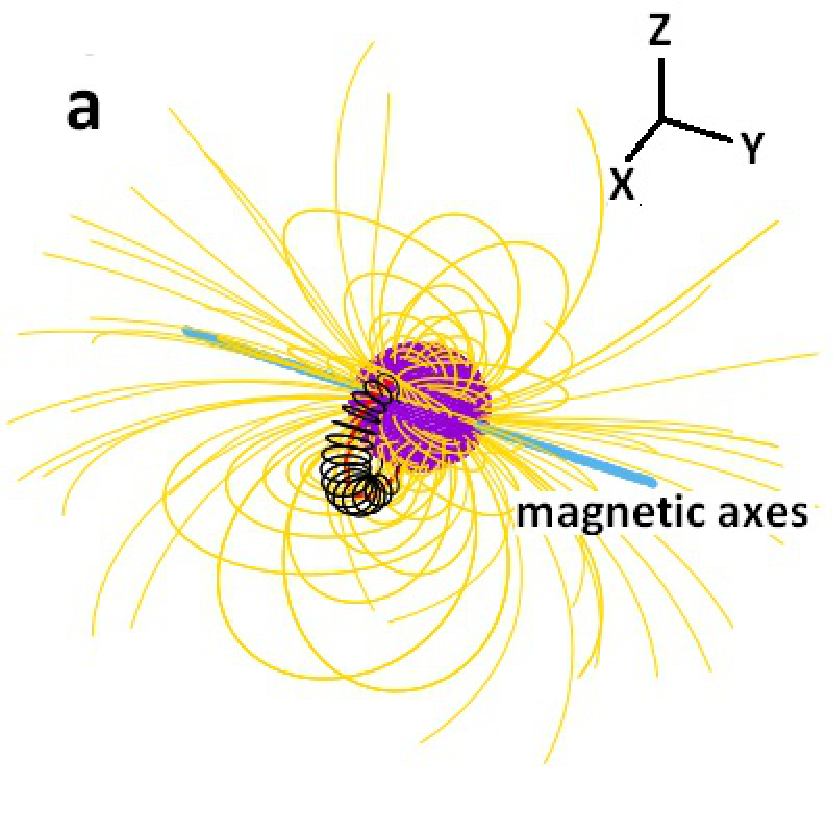}
	\includegraphics[width=1.0\columnwidth]{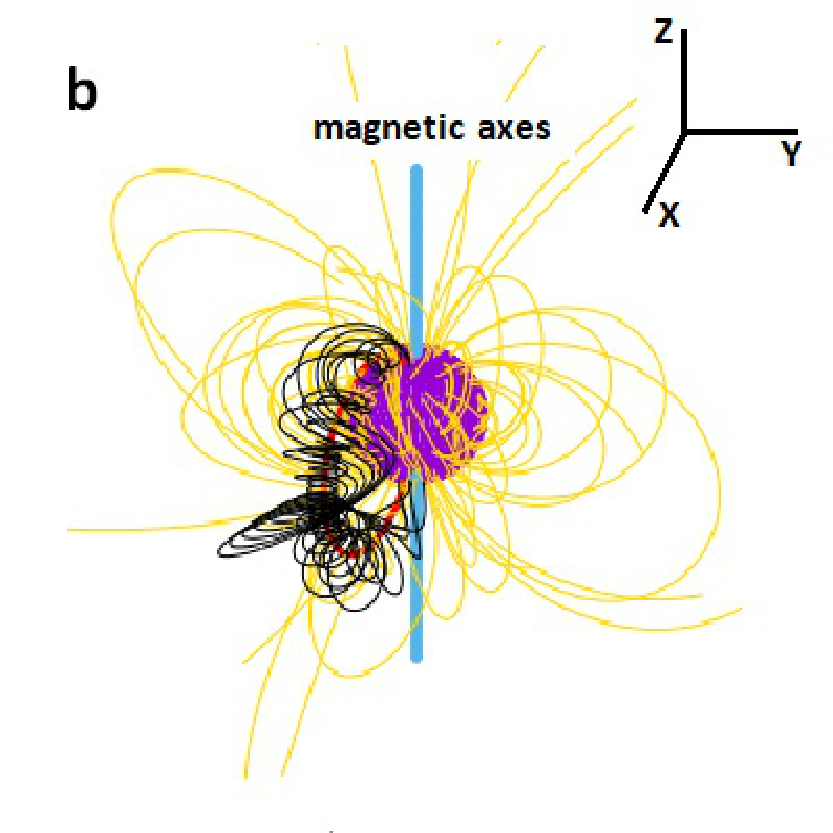}
    \caption{The magnetic field lines near the stellar surface for two special cases. In panel a, the magnetic axes is ${\bf{k'}} = {\bf{j}}$, $a=2$ and $i=0.1$. In panel b, the magnet axes is ${\bf{k'}} = {\bf{k}}$, $a=3$ and $i=0.05$. The dimensionless current $i$ is $i=I/(B_0 c)$. The magnetic axes of the magnetar shown as the cyan lines, yellow curves are for the magnetic field outside the flux rope, black curves are for the field lines near the surface of the flux rope, and red curve describes the axis of the flux rope.}
    \label{blines3d}
\end{figure}

Fig.\ref{blines3d} displays the magnetic configuration given by Eq.(\ref{B0xyz})
is two specific cases with ${\bf{k'}} = {\bf{j}}$ and ${\bf{k'}} = {\bf{k}}$, respectively.
For ${\bf{k'}} = {\bf{j}}$ (Fig.\ref{blines3d}a), the central axis of the flux rope
is located in the zx-plane, perpendicular to the meridian plane. Because of
the symmetry, the magnetic field on the plane $z=0$ does not have z-component,
and the field lines are all located on this plane. Meanwhile, the xy-plane also
poses the top point of the flux rope, and is perpendicular to the central axis
of the flux rope at that point. For ${\bf{k'}} = {\bf{k}}$, the magnetic axis of
the background dipole and the rotating axis are co-located in space, and the flux
rope axis is in the meridian plane. Generally,
the orientation of the flux rope should be somewhere between the two special cases.

Now, we are ready to look into the equilibrium in the flux rope, and then in
the global configuration. Investigating the force acting on the flux rope indicates
that in the case of ignoring the centrifugal force, constraint on the flux
rope that prevents the flux rope from losing equilibrium becomes weakest as
${\bf{k'}} = {\bf{j}}$. In this case, the flux rope is located in the xy-plane,
and only the x-component of the force exists. Studying the equilibrium of the
configuration and its dependence on the relevant parameters in this case. We are
able to deduce the necessary condition for the magnetar that could produce the
giant flare. This condition is also a lower limit to the constraint on the
criterion for the specific magnetars.

To simplify our calculations, we perform non-dimensionalization for the force
and the other parameters, as well as the related equations. Firstly, we define a
dimensionless parameter, $\eta= GMm / (R^{2}B_{0})^{2}$, which relates
the background field, $B_{0}$, to the mass in the flux rope, $m$. Secondarily, the
dimensionless electric current inside the flux rope, $i$, and the dimensionless
field $b_{s}$ and $b_{q,y}$, corresponding to $B_{s}$ and $B_{q,y}$, respectively,
read
\begin{eqnarray} \label{dimlesspara2}
&&i = \frac{I}{{{B_0}c}},\\ \nonumber
&&{b_s} = \frac{c}{I}{B_s} = \frac{1}{a}\left(\ln \frac{{8a}}{{{r_0}}} - 1\right),\\ \nonumber
&&{b_{q,y}} = \frac{c}{I}{B_{q,y}} = {a^2}[2a(x{q_{340}} + z{q_{331}}) - (ax + 1){q_{320}}].
\end{eqnarray}
Therefore, the total force can be written as
\begin{eqnarray} \label{dimlessforceX}
&&{f_x} = {B_0}^2\int\limits_{{\theta _1}}^{{\theta _2}} \{ -\frac{{\eta x}}{{{\theta _2} - {\theta _1}}} \frac{1}{{{r^3}}} \\ \nonumber
&& - a\left[({b_s} + {b_{q,y}}){i^2} - \frac{{u}}{{{r^3}}}i \right]\cos 2\theta \} d\theta.
\end{eqnarray}

\begin{figure}
	\includegraphics[width=1.0\columnwidth]{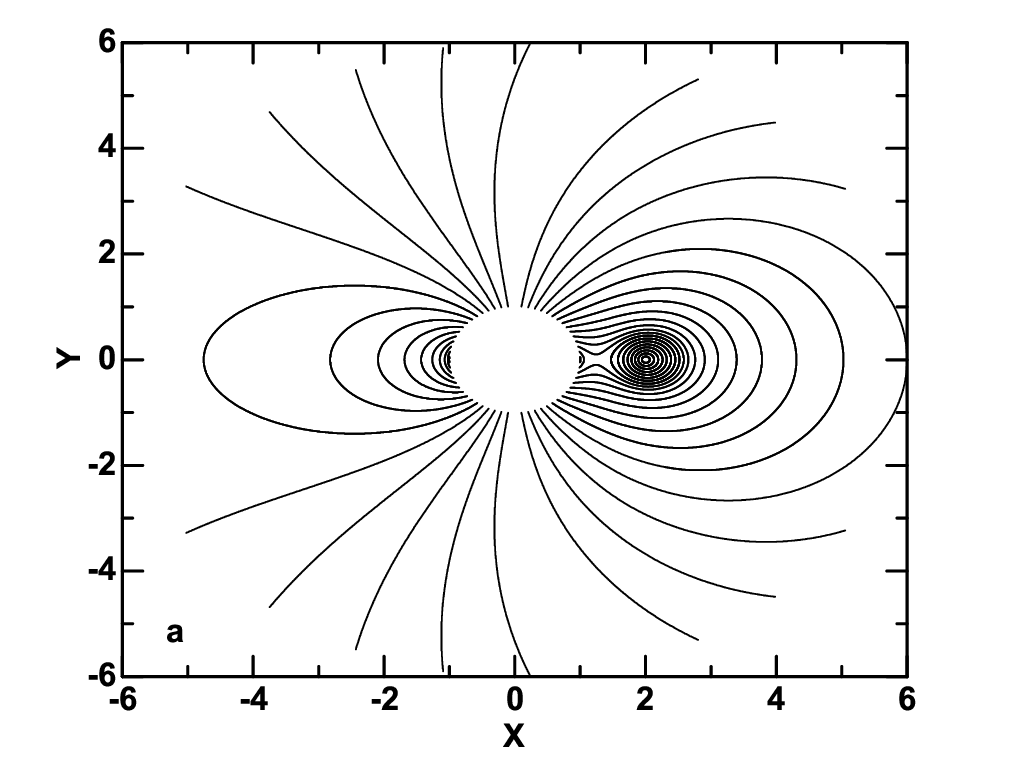}
	\includegraphics[width=1.0\columnwidth]{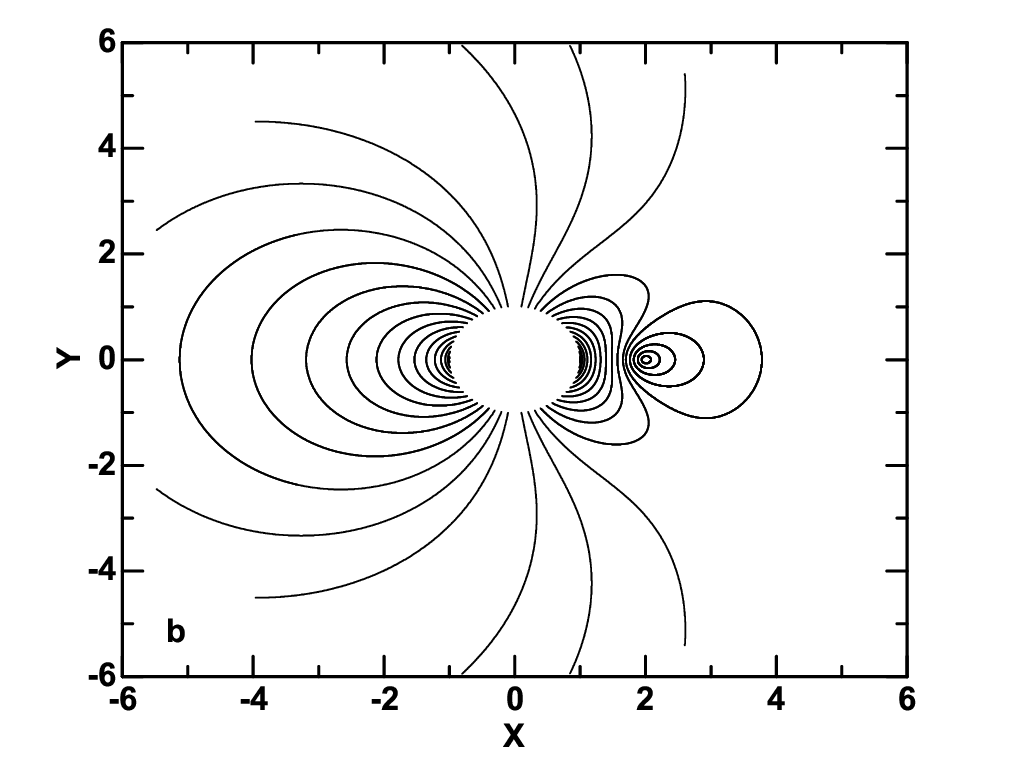}
    \caption{The magnetic field lines at $z=0$ plane for the case of ${\bf{k'}} = {\bf{j}}$, which means the magnetic axes is upward. In panel a, $a=2$ and $i=0.1$. In panel b, $a=2$ and $i=-0.1$. The dimensionless current $i$ is $i=I/(B_0 c)$. A positive $i$ means that the direct of current at the top of the flux rope (i.e., $x=a=2$ and $y=0$) is passing through the paper from the line of sight. Panel a corresponds to the panel a of Fig.\ref{blines3d}.}
    \label{blines2d}
\end{figure}

As mention before, this work focuses on the case of  ${\bf{k'}} = {\bf{j}}$, so
the flux rope is located on the magnetic equator where $B_{0,x}=B_{0,z}=0$ and
thus $f_y=0$. Setting $f_x=0$ determines the equilibrium location of the flux
rope in the magnetosphere,and the relation of the current in the flux rope $I$ to
the other parameters could be established via the frozen-flux condition that
the magnetic field lines on the surface of
the flux rope remain unchanged \citep[e.g., see discussions of][]{FI91,Iea93,IF07}.
For the present case, the frozen-flux condition on the surface of the flux rope is ${\bf{A}}(x=a-r_{0},y=0,z=0)={\bf{const}}$.
Because of the symmetry, ${\bf{A}}=A_{z}{\bf{k}}$ at $(x,y=0,z=0)$, and
the frozen-flux condition on the surface of the flux rope becomes $A_{z}(x=a-r_{0},y=0,z=0)=const$.
Applying this condition to Eq.(\ref{AGreen}), and completing the associated
integrals listed in Eqs.(\ref{AV}) through (\ref{defpq}), we obtain the
the frozen-flux condition is
\begin{eqnarray} \label{frozen}
i [({p_{100}} - {q_{100}}) - 2({p_{120}} - {q_{120}})]\mid_{(a-r_{0},0,0)} - \frac{u}{(a-r_0)^2} = c_1,
\end{eqnarray}
with consideration of the non-dimensionalization described earlier.
Here $(a-r_{0}, 0, 0)$ denotes a point at the surface of the flux rope on the
$x$-axis, and $c_1$ is a parameter that is usually determined by the value of $A_{0,z}$ as
the electric current reaches maximum \citep[e.g., see also][]{Ltl98}. Following
the practice of \citet{Ltl98}, the internal equilibrium gives
\begin{eqnarray} \label{r0i}
r_0 i = c_2,
\end{eqnarray}
where $c_2$ is a parameter related to the size of the rope. Equation (\ref{r0i}) relates the internal
parameter of the flux rope, $r_0$, to the global one $i$. The evolution in the
system is now described by the change in $a$ and current $I$ in response to the
variation of the background field, $u$.
For a given $u$, the equilibrium condition with Eq.(\ref{dimlessforceX}) being zero
and the frozen condition Eq.(\ref{frozen}) determine the height $a$ and the current
$I$ of the flux rope in equilibrium. The gradual change in the background field,
$u$, drives the system in equilibrium to evolve in a quasi-state fashion, and in
this process, $\eta$, $c_1$ and $c_2$ are free parameters.

We note here that the evolutionary fashion of the background field in the magnetar
magnetosphere may not follow the same pattern as what happens in the solar atmosphere.
The crust surface of the magnetar may prevent magnetic field from moving or emerging
on the star surface easily. But the local motions of the broken pieces of the crust
could bring the anchored magnetic field to interact with the field nearby, weakening
or enhancing the magnetic field via magnetic reconnection or merging \citep[e.g., see also discussions of][]{MLZ14}.
This could cause the background field in the magnetar magnetosphere to change.

It is not convenient to
directly set a value for $c_1$ in the present case. Alternatively, we set an initial
height $a=a_0$ as a free parameter that $a_0$ is little larger than unity, and $I$ can
be obtained by using the equilibrium condition with the initial state $u=1$, and
then $c_1$ is determined. Here we take $a_0=1.02$ and $c_2=10^{-5}$. No significant
difference has been found as $a_0$ varies from 1.01 to 1.10 or change $c_2$ by two
orders of magnitude. Finally, parameter $\eta$ is determined by the mass of the rope
and the strength of the background magnetic field.
In our calculation, we solve the equilibrium condition and the frozen condition at
different height $a$ from $a_0$ to $a=3$ to get $I$ and $u$.

\section{Results} \label{SecRes}

\begin{figure}
	\includegraphics[width=1.0\columnwidth]{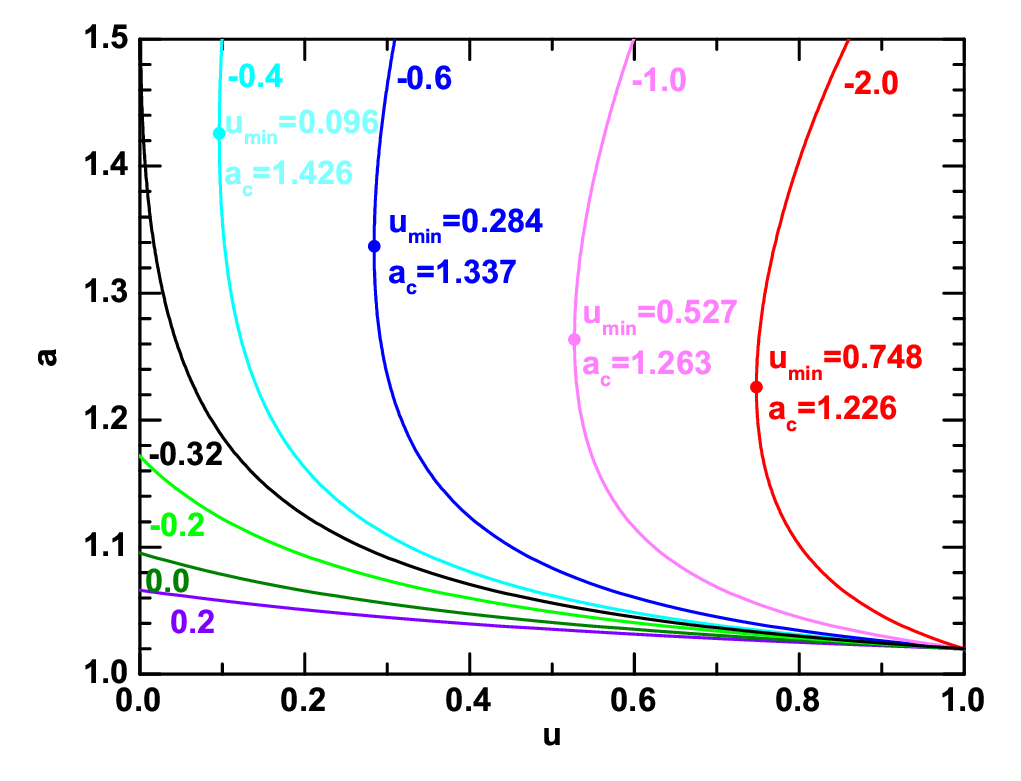}
    \caption{The height of the flux rope equilibrium locations as a function of the relative strength of the background dipole $u$ for different parameter $\eta$.
    The values of $\log \eta$ take $0.2$, $0$, $-0.2$, $-0.32$, $-0.4$, $-0.6$, $-1.0$ and $-2.0$ for each line, they are represented by purple, dark green, light green, black,
    light blue, dark blue, pink, and red colors, respectively. }
    \label{samples}
\end{figure}

The numerical simulation work of \citet{Yu12} shows that flux injection prior to large-scale
eruptions would lead to a slow change in the background magnetic field. In our model, the
evolution of the background magnetic field can also cause the flux rope to lose balance and
produce eruptions. According to the formula derived in the previous section, we can plot curves
showing the variation of the equilibrium height $a$ of the magnetic flux rope with the evolution
of the background magnetic field $u$ for different values of $\log \eta $, as shown in Fig.\ref{samples}.
Each curve in Fig.\ref{samples} can be divided into two regions. In the left region of each curve,
the net force acting on the magnetic flux rope is outward (${f_x}>0$), away from the neutron star.
In the right region of each curve, the net force acting on the magnetic flux rope is inward (${f_x}<0$),
toward the neutron star.

The equilibrium curves in Fig.\ref{samples} can be classified into two types: those with nosepoints
and those without. For equilibrium curves without nosepoints, the equilibrium height $a$ of the
magnetic flux rope monotonically increases with the decrease in the background magnetic field $u$.
This implies that regardless of how the intensity of the background magnetic field changes, catastrophic
behavior will not occur, and there is no possibility of losing equilibrium and causing an eruption
in the system. In contrast, for equilibrium curves with nosepoints, as the background magnetic field $u$
decreases, the equilibrium height $a$ of the magnetic flux rope gradually increases until it reaches
a critical point. At or beyond this critical point, the magnetic flux rope cannot maintain stable
equilibrium and undergoes an eruption.

Referring to Fig.\ref{samples}, for equilibrium curves with $\log \eta \leq -0.32$, the decay of the
background magnetic field $u$ can lead the system to evolve to the critical height $a_c$ of the magnetic
flux rope. Taking $\log \eta = - 2.0 $ as an example, the critical value of the background magnetic
field $u_c$ is 0.748, and the critical height of the magnetic flux rope $a_c$ is 1.226. The existence
of a minimum in the evolution of the background magnetic field $u$ indicates that when $u$ decreases
below this minimum value, there is no longer a height at which the magnetic flux rope can maintain
a stable equilibrium. Our computational results suggest that when the background magnetic field $u$
evolves to the critical value, $u<u_{\min}$, the net external force acting on the magnetic flux rope
points outward from the star. At this point, the magnetic flux rope is accelerated away from the stellar
surface under the influence of the net external force, triggering the catastrophic event leading to
an eruption.

\begin{figure}
	\includegraphics[width=1.0\columnwidth]{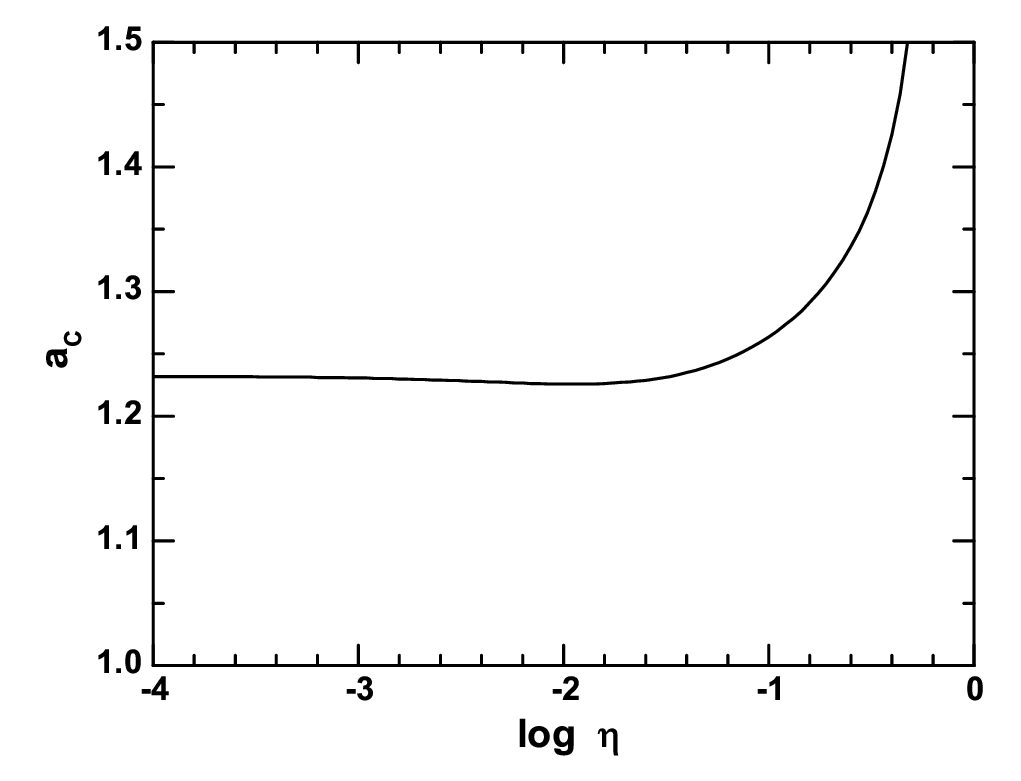}
    \caption{The critical height as a function of parameter $\eta$. }
    \label{etaac}
\end{figure}

Fig.\ref{etaac} shows the critical height values $a_c$ of equilibrium curves obtained through
calculations for different values of $\log \eta $. From Fig.\ref{etaac}, it can be seen that when
$\log \eta < -0.32$ or $\eta < 0.48$, the equilibrium curves of the system evolution exhibit
nosepoints/critical values. For a neutron star with a radius $R=10^6$cm and mass $M=3\times10^{33}$g,
substituting $\eta < 0.48$ into equation $\eta = G M m / ({{{R^4}{B_0}^2}}) \approx 200 (m/{\rm g}) (B_0/{\rm G})^{-2}$, yields:
\begin{eqnarray} \label{critical}
\left(\frac{B_{0}}{\rm G}\right)^{2} > 417 \left(\frac{m}{\rm g}\right),
\end{eqnarray}
where g is gram and G is Gauss.
For neutron stars,
there is a well-known relation between the magnetic field, the rotational period $P$ and its
derivative $\dot{P}=dP/dt$:
\begin{eqnarray} \label{BPP}
{B_0}^2 \approx \frac{{3M{c^3}P \dot{P}}}{{20{\pi ^2}{R^4}}}.
\end{eqnarray}
By combining Eq.(\ref{critical}) and (\ref{BPP}), we obtain
\begin{eqnarray} \label{PPm}
\left(\frac{P\dot{P}}{\rm s}\right) > 3.3\times 10^{-33} \left(\frac{m}{\rm g}\right),
\end{eqnarray}
where s is second. The Australia Telescope Compact Array and the Very Large Array observed an expanding radio nebula
associated with the giant flare from magnetar SGR 1806-20 \citep{Tay05,Cam05,Gae05},
which implies a large scale mass ejection in the eruptive process that produced the giant flare.
If the ejecta is roughly spherical, the radiation is well-explained by synchrotron
emission from a shocked baryonic shell with a mass of $\geq 10^{24.5}$~g \citep{Gae05,Gra06,Mas10}.

We have calculated the light curves of the eruption with $10^{25}$~g and $10^{26}$~g,
respectively, and the resultant light curves are in good agreement with observations\citep{MLZ14}.
Therefore, we argue that it is reasonable to choose the observation value $m= 10^{24.5}$~g given
by ATCA and VLA as a lower limit to the true mass ejected during the giant flare.
So for $m \sim 10^{24.5}$~g, Eq.(\ref{critical}) gives $B_{0} > 3.6\times 10^{13}$~G, and Eq.(\ref{PPm})
gives $P \dot{P} > 1.1 \times 10^{-12}$~s.

\begin{figure}
	\includegraphics[width=1.0\columnwidth]{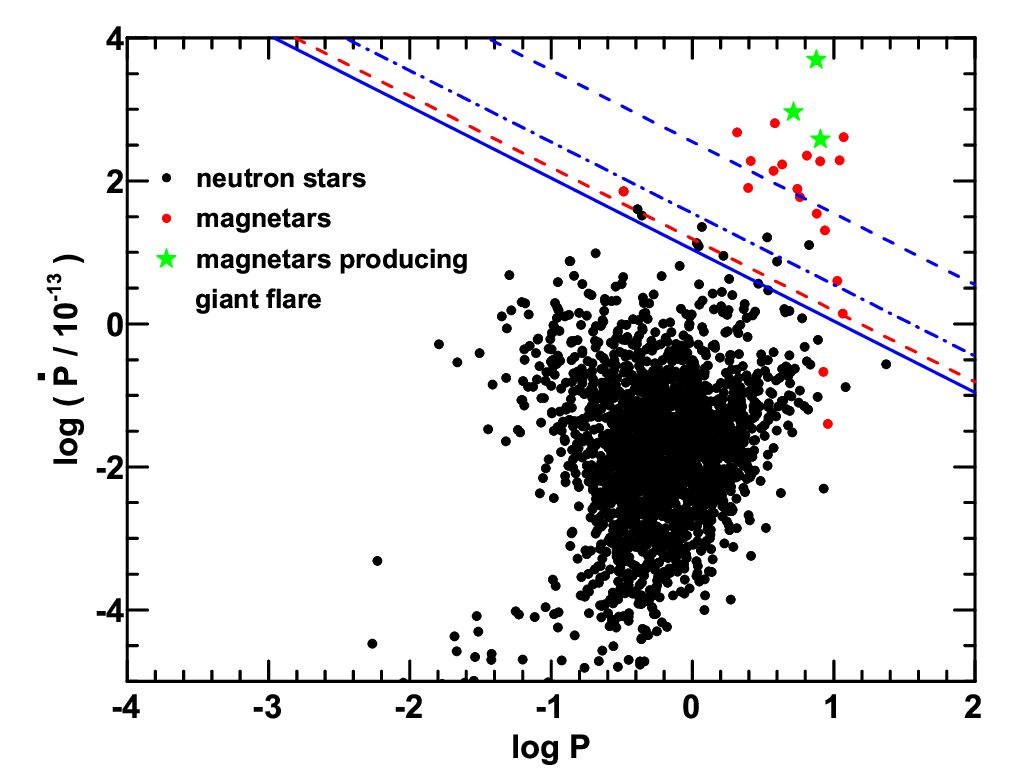}
    \caption{The $P-\dot{P}$ distribution of neutron stars. Magnetars that have produced giant flares are denoted as green stars,
    the other magnetars are denoted as the red dots, the normal neutron stars are denoted as the black dots.
    The data of magnetars are available from McGill Online Magnetar Catalog \citep{OK14} on the website: (http://www.physics.mcgill.ca/~pulsar/magnetar/main.html).
    The data of pulsars are available from ATNF Pulsar Catalogue \citep{Man05} on the web address:(http://www.atnf.csiro.au/research/pulsar/psrcat).
    The blue solid line corresponds to $B_{0} = 3.6\times 10^{13}$G for ejected mass of $10^{24.5}$g (see Eq.(\ref{critical})) obtained by our model,
    the blue dot-dashed line corresponds to $B_{0} = 6.5\times 10^{13}$G for ejected mass of $10^{25}$g,
    and the blue dashed line corresponds to $B_{0} = 2\times 10^{14}$G for ejected mass of $10^{26}$g.
    The red dashed line corresponds to the quantum electron critical magnetic field, $B_{crit}$, for comparison.}
    \label{pdp}
\end{figure}

The relationship of $P\dot{P} = 1.1 \times 10^{-12}$~s plotted
in the diagram of the $P - \dot{P}$ distribution for pulsars as the blue solid line in Fig.\ref{pdp}.
This determines a lower limit to the energetics of the giant flare corresponding to the lower limit to
the ejected mass from magnetars. For different ejected masses, the relationships in Eq.(\ref{critical})
and Eq.(\ref{PPm}) gives the other lines in Fig.\ref{pdp}:  the blue dot-dashed line corresponds
to $B_{0} = 6.5\times 10^{13}$~G for ejected mass of $10^{25}$~g, and the blue dashed line corresponds
to $B_{0} = 2\times 10^{14}$~G for ejected mass of $10^{26}$~g, which outlines a region in the parameter
space for the magnetars that could produce giant flares associated with ejections of large mass.
The red dashed line corresponds to the quantum electron critical magnetic field,
$B_{crit}=m_{e}^{2}c^{3}/(e \hbar) = 4.4\times 10^{13}$~G, for comparison. Here $m_{e}$ is the electron
mass, $e$ is the electron charge, and $\hbar$ is the Plank constant.

\section{Discussions} \label{SecDis}

In this work, we developed a catastrophe model
of the giant flare from magnetars following the practice of \citet{MLZ14}.
 The model is constructed in the spherical coordinate system without any symmetry,
 but the magnetar has a spherical surface.
 We considered the physical process of quasi-static magnetic energy accumulation that
 causes the catastrophic loss of equilibrium of a twisted flux rope in the magnetar's magnetosphere.
 This model establishes criteria for identifying magnetars capable of producing giant flares.

\subsection{Brief Overview of the Model Construction}\label{SecBrief}
The model incorporates a magnetic field configuration closer to reality, where the two footpoints
of the magnetic flux rope are on the surface of the neutron star. Therefore, axial symmetry in
the magnetic configuration studied by \citet{Ltl98} no longer exists, making the corresponding
calculations more complex. In our calculations, we assume that the flux rope is initially located
on equatorial plane, and is a part of a circle passing the stellar center. In this way,
the calculation is simplified and it is easy to derive analytical expressions. We focus
on the necessary conditions for the instability of the magnetic flux rope. The occurrence of
instability is most favorable when the magnetic axis is perpendicular to the plane of the
magnetic flux rope.

 During the quasi-static evolution of the magnetic flux rope, it is subject to the forces of
 magnetic, gravitational, centrifugal, and curvature forces. When these forces balance each
 other, the magnetic flux rope is in equilibrium. Within a certain range, despite changes
 in the background magnetic field, the overall magnetic configuration can maintain equilibrium
 through self-regulation, and the system remains in a stable equilibrium state, evolving slowly
 in a quasi-static manner. As the background magnetic field changes, magnetic energy also slowly
 and continuously accumulates in the system until it reaches a threshold. At this point, the
 system can no longer maintain equilibrium through self-regulation, and further evolution
 leads to the loss of equilibrium and the occurrence of an eruption.

 We find that in this
 process, the centrifugal force is an order of magnitude smaller than gravity, so its influence
 is temporarily neglected in this work. The overall magnetic field of the magnetar dipole may
 not change for a reasonably long time; but in a relatively local region near the surface of
 the magntar, various interactions of the complex magnetic structure could occur. The interaction
 could be either reconnection or merging among magnetic structures, which leads to change in
 the magnetic field around the area of interest.

 We describe the evolution of the system by varying the parameter $\eta$ in the calculations.
 Based on the equilibrium conditions of different background magnetic fields and the
 frozen-flux condition on the surface of the flux rope, we obtain a set of equilibrium
 curves (see Fig.\ref{samples}). Depending on whether a turning point exists, the equilibrium
 curves are divided into two types. The first type, without a critical point, exhibits a
 monotonic and slow increase in height with the decay of the background magnetic field. In this
 case, catastrophic behavior does not occur. The second type of the equilibrium curve has the weakest
 background magnetic field at the critical of the equilibrium curve. This means that
 when the evolution of the background magnetic field intensity reaches this minimum value,
 the quasi-static evolution in the system terminates, and the flux rope is ejected away from the neutron star.
 Therefore, evolution described by the second type of equilibrium curve can lead to large-scale
 magnetic energy release, which can be used to study rapid magnetic energy release processes in
 celestial bodies.

\subsection{Key Results}\label{SecKey}
 Based on the characteristics of the second type of the equilibrium curve, we further deduced the
 relationship between the critical height $a_c$ and the parameter $\eta$, determining the range of
 values for $\eta$ when a critical point exists during the evolution in the system ($\eta < 0.48$).
 By substituting the characteristic
 parameters of magnetars into the relationship, we find that the necessary condition for a catastrophic
 eruption in a magnetar is $B_0 > 3.6 \times 10^{13}$G, or $\eta < 0.48$. According to this result, we draw a
 blue solid line on the diagram of all known isolated pulsars, dividing pulsars in Fig.\ref{pdp}
 into two parts. The lower-left part below the blue solid line includes most normal radio pulsars with
 magnetic fields weaker than $3.6 \times 10^{13}$G, while the upper-right part above the blue solid line contains fewer
 stars, mostly magnetars.

 According to our model, pulsars in the region below the blue solid line are
 not expected to undergo catastrophic eruptions, while those in the region above the blue solid line
 could store sufficient magnetic energy, leading to the loss of equilibrium in the magnetic
 configuration and the occurrence of giant flares.  Currently, three magnetars have been reported
 to produce giant flares, and they are denoted by green star
 above the blue solid line in Fig.\ref{pdp}, indicating that our results are consistent with observations.

 Within this framework, \citet{MLZ14} developed a theoretical model for giant flares from magnetars, and our
 work provides a criterion for predicting which pulsars would produce giant flares. The red dashed
 line in Fig.\ref{pdp} represents the quantum electron critical magnetic field
 $B_{crit} = m_{e}^{2} c^{3} / e\hbar \sim 4.4\times 10^{13}$G, where the cyclotron energy of an
 electron reaches the electron rest mass energy. This line distinguishes strongly magnetized
 neutron stars from ordinary neutron stars \citep{RE11}, and the magnetic field required for pulsars
 capable of producing giant flares is a little lower than this critical magnetic field.

In addition to neutron stars, we also use our model to evaluate the mass of ejection of the solar CME.
 Masses of CMEs deduced from SOLWIND observations in the time interval between 1979 to 1981 \citep{Ho85}
 range from $2\times 10^{14}$~g to $4\times 10^{16}$~g with an average of $4.1\times 10^{15}$~g \citep[e.g., see
 also discussions of][]{Lin04}. Substituting the relevant parameters of the Sun
 into the eruption conditions derived in this paper, we find that the maximum ejected mass of solar CMEs
 does not exceed $10^{17}$g, which is consistent with all current observations.

\citet{Ste23} demonstrated that locally entangled and braided dipole fields in magnetars can lead to
small/mid-scale eruptions through rapid magnetic reconnection, which prevents significant energy buildup
and subsequent giant flares. Their work showed complex magnetic structures readily form in magnetar
magnetospheres, while giant flares are statistically rare though not impossible.

On the basis of their foundation, our work examines how such flux-rope configurations may catastrophically
lose equilibrium in response to the evolution in the background field, potentially triggering giant flares (Fig.\ref{pdp}).
Thus we confirm that while energetically challenging, giant flares can indeed occur through equilibrium
collapse in these systems, complementing findings by \citet{Ste23} on structure formation with mechanistic
insight into the eruption physics.

\subsection{Limitations} \label{SecLim}

In this work, we are focusing on whether a lower limit to the production of giant flares from magnetars could be established on the basis of simple model. The ejecta associated with the giant flare in the model is an arc-shaped flux rope carrying electric current initially embedded in the dipolar background magnetic field of the magnetar. The electric current flows through the flux rope that fills with plasma.

In the solar case, the plasma inside the flux rope consists of protons and electrons mainly, and the total mass is basically determined by the proton. In the case of magnetar, on the other hand, it should be contributed from all particles if the material inside the flux rope is the pair-plasma. In principle, the magnetic field plays the role in the formation of the rope, including the internal structures, and may impact the total mass input to the flux rope, but may not govern the relevant mass composition.

For the time being, limited by the observational/detecting technique, we do not have definite knowledge about the total mass and the mass composition of the flux rope involved in the eruption from magnetars. So the total mass is used as a tunable parameter in the related studies. As for the sensitivity of the total mass to the composition, we believe that it should be related to the composition indeed, but whether the former is very sensitive to the latter is in fact beyond the scope of our knowledge. According to the electromagnetic property of plasma and also to the knowledge about the solar eruption, no matter whether it is hydrogen plasma or pair-plasma, it should be the total mass that impacts the evolution of the system, and the composite of the plasma may not play any role in the process. To our knowledge, the composite of the plasma might impose constrains on the other issues studied in different works, but may not show its impact on the issues studied in the present work.

Here we note that the relationship between flux rope size ($r_{0}$) and current intensity ($I$)
adopted in this study is based on one possible current distribution configuration within the
flux rope. According to \citet{Iea93} and \citet{Rug24}, several types of electric current distributions
inside the flux rope could exist, and different current distributions can affect the local equilibrium in different fashions.
Due to limited observational constraints, we employ a simplified Lundquist force-free field
solution, where the magnetic twist and current profiles follow Bessel functions,
$J_{0}$ and $J_{1}$ \citep[e.g., see also][for details]{Ltl98}. Future work should incorporate
more realistic distributions when additional internal structure data become available.

\section{Conclusions} \label{SecCon}

 This study builds upon the giant flare theory model for magnetars developed by \citet{MLZ14} and
 utilizes the mathematical methods of \citet{Ltl98} to develop a three-dimensional theoretical model
 for giant flares in magnetars. We considered the physical process that causes the catastrophic loss
 of equilibrium of a twisted flux rope in the magnetar magnetosphere. The loss of equilibrium behavior
 of a twisted flux rope is investigated in a star with dipolar magnetic field, taking into account
 possible effects of the changes in background magnetic field.

 In this work, we use a three-dimensional
 flux rope to describe a magnetic arch in the magnetar's magnetosphere. When all the forces acting on the
 magnetic arch balance one another, it remains in a stable equilibrium state for a certain period, evolving
 slowly in a quasi-static manner in response to the gradual change in the background magnetic field. Once a critical
 point is reached, the stable equilibrium cannot be maintained, leading to a transition to the dynamic evolution,
 triggering the catastrophe and causing the eruption. By examining the equilibrium configurations
 under different conditions and the evolution process leading to critical states, we obtain the range of
 parameters required for a neutron star to produce the giant flare.

 Utilizing these results,
 and based on the relationship among magnetic field, period, and period derivative of pulsars, we
 distinguish pulsars that can produce giant flares in the $P - \dot{P}$ diagram of all identified isolated pulsars.
 Therefore, our model provides a reasonable interpretation of the mechanism for the giant
 flare from magnetars, and set up a criterion for predicting which kind of pulsars could produce giant flares.
 In the future, we will further refine the model and apply it to investigating and exploring similar intense
 magnetic active phenomena occurring in both magnetars and the other environments.

\section*{Acknowledgements}

We appreciate the referee for valuable comments and suggestions that have helped improve this work apparently. This work was supported by the National Key R\&D Program of China (grant Nos. 2022YFF0503804, 2021YFA1600400/2021YFA1600402), the National Natural Science Foundation of China (grant Nos. 12288102, 11773064, 12273107), the Strategic Priority Research Program of the Chinese Academy of Sciences (grant No. XDB 41000000), the Foundation of the Chinese Academy of Sciences (Light of West China Program and Youth Innovation Promotion Association), the Yunnan Province Scientist Workshop of Solar Physics, the Yunnan Revitalization Talents Support Program, the International Centre of Supernovae, Yunnan Key Laboratory (No. 202302AN360001), and the Yunnan Key Laboratory of Solar Physics and Space Science (202205AG070009). JL benefited from the discussions of the ISSI-BJ Team ``Solar eruptions: preparing for the next generation multi-wave band coronagraphs.'' The numerical computation in this paper was carried out on the computing facilities of the Computational Solar Physics Laboratory of Yunnan Observatories (CosPLYO).

\section*{Data Availability}

The data underlying this article will be shared on reasonable request to the corresponding author.












\appendix

\section{ The method of Green's function for solving Poisson's equation }\label{AppA}

The method of Green's function for solving Poisson's equation can be found in many books
of mathematical physics \citep[e.g.,][]{Has13}. Here we summarize the basic approach. For the Poisson's
equation ${\nabla ^2}\varphi ({\bf r}) =  - \rho ({\bf r})$, considering a Green's function
satisfying ${\nabla ^2}{\widetilde G}({\bf r},{\bf r}') =  - \delta ({\bf r} - {\bf r}')$, there is
\begin{eqnarray} \label{ApdxA1}
&& \oint {(\varphi {\nabla {\widetilde G}} - {\widetilde G} {\nabla \varphi })d{\bf S} } = \int {(\varphi {\nabla ^2}{\widetilde G} - {\widetilde G}{\nabla ^2}\varphi )dV}  \\ \nonumber
&& = \int {[ - \delta ({\bf r} - {\bf r}')\varphi  + {\widetilde G}\rho ({\bf r})]dV} =  - \varphi ({\bf r}') + \int {{\widetilde G}\rho ({\bf r})dV},
\end{eqnarray}
where the domain of integral covers the source region denoted by ${\bf r}'$.
Therefore we have
\begin{eqnarray} \label{ApdxA2}
\varphi ({\bf r}') = \int {{\widetilde G}\rho ({\bf r})dV}  + \oint {({\widetilde G}{\nabla \varphi } - \varphi {\nabla {\widetilde G}})d{\bf S} }.
\end{eqnarray}
Since ${\widetilde G}({{\bf r}},{{\bf r}}') = {\widetilde G}({{\bf r}}',{{\bf r}})$, exchanging ${{\bf r}}$ and ${{\bf r}}'$ yields
\begin{eqnarray} \label{ApdxA3}
\varphi ({\bf r}) = \int {{\widetilde G}\rho ({\bf r}')dV'}  + \oint {({\widetilde G}{\nabla' \varphi } - \varphi {\nabla' {\widetilde G}})d{\bf S}' },
\end{eqnarray}
where the domain of integral covers the field region denoted by ${\bf r}$. Eq.(\ref{ApdxA3}) is the general solution of the Poisson's equation.

When boundary condition of $\varphi$ is supplemented, the corresponding boundary condition of Green's function ${\widetilde G}$ is required to determine ${\widetilde G}$, i.e., the general boundary condition for the linear Poisson's equation is that the value of $a\varphi+b(\partial \varphi / \partial n)$ on a closed surface are given, where $a$ and $b$ here are known function of ${\bf r}$, the corresponding boundary condition of Green's function is $a{\widetilde G}+b(\partial {\widetilde G} / \partial n)=0$ on the closed surface. Usually, this closed surface is chosen to be boundary of the given configuration. For the problem we are dealing with here, $\widetilde G$ vanishes for ${\bf r}'$ on the boundary, and $b=0$.

\section{ Calculation of the integrals $p_{smn}$ }\label{AppB}

The purpose of this part of work is to demonstrate how the integral below is closed:
\begin{eqnarray} \label{Apdxp}
&&{p_{smn}} := \int\limits_{{\theta _1}}^{{\theta _2}} {g{{_1}^s}{{\sin }^m}\theta '{{\cos }^n}\theta 'd\theta '},
\end{eqnarray}
where
\begin{eqnarray} \label{Apdxg}
g_1 = \frac{1}{{\sqrt {{r^2} + {a^2}{{\sin }^2}\theta ' - 2ar\sin \theta '\cos \gamma } }}, \\ \nonumber
\cos \gamma  = \cos \theta \cos \theta ' + \cos \phi \sin \theta \sin \theta '.
\end{eqnarray}

We first define some symbols that we are going to use for the calculation (only used in this appendix):
\begin{eqnarray} \label{Apddef}
&&A := {r^2} - ar\cos \phi \sin \theta  + \frac{1}{2}{a^2} \\ \nonumber
&&B := - a\left(\frac{a}{2} - r\cos \phi \sin \theta \right) \\ \nonumber
&&C := - ar\cos \theta  \\ \nonumber
&&{\theta _0} := \frac{1}{2}{\rm{sign}}(C){\cos ^{ - 1}}\frac{B}{{\sqrt {{B^2} + {C^2}} }} \\ \nonumber
&&{k^2} := \frac{{2\sqrt {{B^2} + {C^2}} }}{{A + \sqrt {{B^2} + {C^2}} }},
\end{eqnarray}
then $g_1$ satisfies:
\begin{eqnarray}\label{Apddef}
g_1^{ - 2} = (A + \sqrt {{B^2} + {C^2}} )[1 - {k^2}{\sin ^2}(\theta ' - {\theta _0})].
\end{eqnarray}

We then define other symbols (also only used in this appendix):
\begin{eqnarray} \label{Apddef2}
&&u:= \sin {\theta _0} \\ \nonumber
&&v:= \cos {\theta _0} \\ \nonumber
&&{w_{s,l,m}}:= \int\limits_{{\theta _1} - {\theta _0}}^{{\theta _2} - {\theta _0}} {\frac{{{{\sin }^l}t{{\cos }^{m - l}}t}}{{{{(1 - {k^2}{{\sin }^2}t)}^{\frac{s}{2}}}}}dt}.
\end{eqnarray}

Using the binomial expansion, $p_{smn}$ can be represented as:
\begin{eqnarray}\label{Apdpsmn}
&&{p_{smn}} = {(A + \sqrt {{B^2} + {C^2}} )^{ - \frac{s}{2}}} \\ \nonumber
&&\sum\limits_{l = 0}^{m + n} {\sum\limits_{j = \max (0,l - m)}^{\min (l,n)} {} } [( - 1)^j}C_m^{l - j}C_n^j{v^{n + l - 2j}}{u^{m - l + 2j}}{w_{s,l,m + n}].
\end{eqnarray}
The integral of $w_{s,l,m}$ is thus expressed analytically as elliptic integrals, and so is $p_{smn}$.


\bsp	
\label{lastpage}
\end{document}